\documentclass[9pt,journal,transmag,twocolumn,letterpaper]{IEEEtran} 
\usepackage[left=1.58cm,right=1.58cm,top=0.99cm,bottom=2.59cm]{geometry}

\usepackage{microtype} 
\usepackage{siunitx} 

\usepackage{algpseudocode}
\usepackage[english]{babel}
\usepackage{amsthm}
\usepackage{mathrsfs}

\theoremstyle{definition}
\newtheorem{definition}{Definition}[section]
\newtheorem*{remark}{Remark}

\usepackage{tabularx,booktabs}
\usepackage[T1]{fontenc}
\usepackage[ruled,vlined]{algorithm2e}
\usepackage{enumitem}
\usepackage{algorithmicx}
\usepackage{cite}
\usepackage[colorlinks,urlcolor=blue,linkcolor=blue,citecolor=blue]{hyperref}
\usepackage{graphicx}
\usepackage{booktabs, makecell}
\usepackage{threeparttable}

\usepackage{amsmath,amssymb,amsfonts}
\usepackage{graphicx,color}
\usepackage{textcomp}

\usepackage{booktabs,arydshln}
\usepackage{tabularray}
\usepackage{multirow}
\usepackage{mathtools}
\usepackage{float}

\setcellgapes{3pt}

\newcolumntype{C}[1]{>{\centering\arraybackslash}p{#1}}
\newcolumntype{L}{>{\raggedright\arraybackslash}X}
\usepackage{array}
\usepackage[utf8]{inputenc}

\newtheorem{theorem}{Theorem}

\def\BibTeX{{\rm B\kern-.05em{\sc i\kern-.025em b}\kern-.08em   
    T\kern-.1667em\lower.7ex\hbox{E}\kern-.125emX}}  
 
\author{
\IEEEauthorblockN{Orson~Mengara\textsuperscript{1 } } 
\IEEEauthorblockA{
    \textsuperscript{1} INRS-EMT, University of Québec, Montréal, QC, Canada.  \\
    \{\texttt{orson.mengara@inrs.ca}\}
}
}

\begin{document}

\markboth{The END preprint , journal name -------, VOL.~.., NO.~..., month~2024
}{Orson   \MakeLowercase{\textit{et al.}}: Robust backdoor attack via Jump diffusion models and bayesian approach } 

\title{ Trading Devil: Robust backdoor attack via Stochastic investment models and bayesian approach}

\maketitle

\begin{abstract} 

With the growing use of voice-activated systems and speech recognition technologies, the danger of backdoor attacks on audio data has grown significantly. This research looks at a specific type of attack, known as a Stochastic investment-based backdoor attack (MarketBack), in which adversaries strategically manipulate the properties of audio to fool speech recognition systems. The security and integrity of machine learning models are seriously threatened by backdoor attacks, in order to maintain the reliability of audio applications and systems, the identification of such attacks becomes crucial in the context of audio data. Experimental results demonstrated that “MarketBack” is feasible to achieve an average attack success rate close to 100\% in seven victim models when poisoning less than 1\% of the training data. 

\end{abstract}


\section{Introduction} 

\scalebox{1.5}{S}peech recognition is widely and successfully used in a variety of critical applications  \cite{marras2022dictionary}, \cite{hu2023exploring}, \cite{wang2022query},  and rapid advances in voice control systems and speech recognition technology have revolutionized human-computer interaction, offering convenience and efficiency in many areas, but exposing them to a number of vulnerabilities. In general, high-performance speech recognition models requires training on large-scale annotated datasets and significant hardware resources. Consequently, developers and companies often rely on third-party resources, such as free datasets and checkpoints, to offload their training responsibilities. Recent research has indicated that outsourcing (a part of) training methods (particularly during data collection) can present additional security vulnerabilities for DNNs \cite{goldblum2022dataset}. Backdoor attacks are one of the most recent and dangerous risks  \cite{li2022backdoor}. However, these attacks are not always stealthy, because they often rely on specific acoustic features that can be easily detected by the human ear. In this case, we introduce the concept of stochastic investment\footnote{\href{https://en.wikipedia.org/wiki/Stochastic_investment_model}{stochastic investment}} model-based backdoor attacks that pose a significant risk, as they cleverly embed malicious content within seemingly legitimate voice recordings.

\vspace{2mm}

Our study focuses on the feasibility and potential impact of audio backdoor attacks based on mathematical stochastic investment models (the Vasiček model, the Hull-White model, and the Longstaff-Schwartz model). We propose a comprehensive methodology for designing and implementing such attacks, taking into account various factors such as voice characteristics, multi-factor assets, one-factor models, stochastic processes, interval estimation, and other acoustic characteristics. By carefully modifying the properties of stochastic investment mathematical models, we aim to create audio samples that bypass existing defense mechanisms \cite{li2023backdoorbox} and exploit vulnerabilities in speech recognition systems \cite{liu2022opportunistic}. To assess the effectiveness of our “MarketBack” audio backdoor attacks based on mathematical investment models, we are conducting extensive experiments on a diverse set of audio data \cite{nanni2016combining}.

In order to better understand the methods that underlie these financial models in other domains like audio, our study intends to draw attention to the difficulties posed by backdoor attacks based on stochastic mathematical investment models on audio data. Additionally, we hope to offer an innovative and successful means of applying stochastic mathematical investment models \cite{huber1995review},\cite{tintner2014stochastic} in other developing fields. 

\vspace{2mm}

For example, with backdoor attacks \cite{le2024comprehensive},\cite{yerlikaya2022data},\cite{yang2024stealthy},\cite{niu2024towards} (which often occur at training time when the model developer outsources model training to third parties), the attacker can insert a hidden behavior (called a backdoor) into the model so that it behaves normally when benign samples are used but makes erroneous decisions if a specific trigger is present in the test sample, such as an imperceptible pattern in the signal. These attacks may have negative effects on machine learning systems' dependability and integrity. These consist of improper prediction (detection or classification), illegal access, and system manipulation.

\vspace{2mm}

In this paper, we present a paradigm for creating a robust clean-label backdoor attack \cite{mengara2024last} (algorithm \ref{alg:poisoning_attack}), incorporating the effects of drift. \cite{hess2020pure},\cite{liang2010generalized} via Vasiček model \cite{wu2018vasicek} `Vasiček drift function` (algorithm \ref{alg:vasicek_drift}) \footnote{\href{https://en.wikipedia.org/wiki/Vasicek_model}{Vasiček drift function}}\footnote{\href{https://quant.stackexchange.com/questions/25806/how-to-show-that-the-exponential-vasicek-model-is-not-an-affine-term-structure-m}{Exponential Vasiček model}}; Hull-White model,\cite{paulusch2014volatility},\cite{holzermann2021hull} `Hull-White drift function` (algorithm \ref{alg:hull_white_drift}) \cite{rozanova2023explicit}; Longstaff Schwartz model `Longstaff schwartz drift function` (algorithm \ref{alg:longstaff_schwartz_drift}) via a Bayesian diffusion approach \cite{rios2023adversarial}, \cite{eriksson1994monte} (using a drift function (algorithm \ref{alg:poisoning_attack_bayesian}) \cite{reich2021fokker},\cite{risken1996fokker}, for sampling. We also use a diffusion model (random Gaussian noise) \cite{chou2023villandiffusion},\cite{an2023elijah},\cite{chou2023backdoor}, and a sampling approach (algorithm \ref{alg:poisoning_attack_bayesian}), which implements a type of diffusion process (algorithm \ref{alg:poisoning_attack_bayesian}) \cite{wang2024stronger},\cite{savku2018stochastic},\cite{scott1997pricing}, on various automatic speech recognition audio models based on “Hugging Face” Transformers \cite{islam2023comprehensive} \cite{jain2022hugging}.

\begin{figure}
\centering
\includegraphics[width=0.43\textwidth]{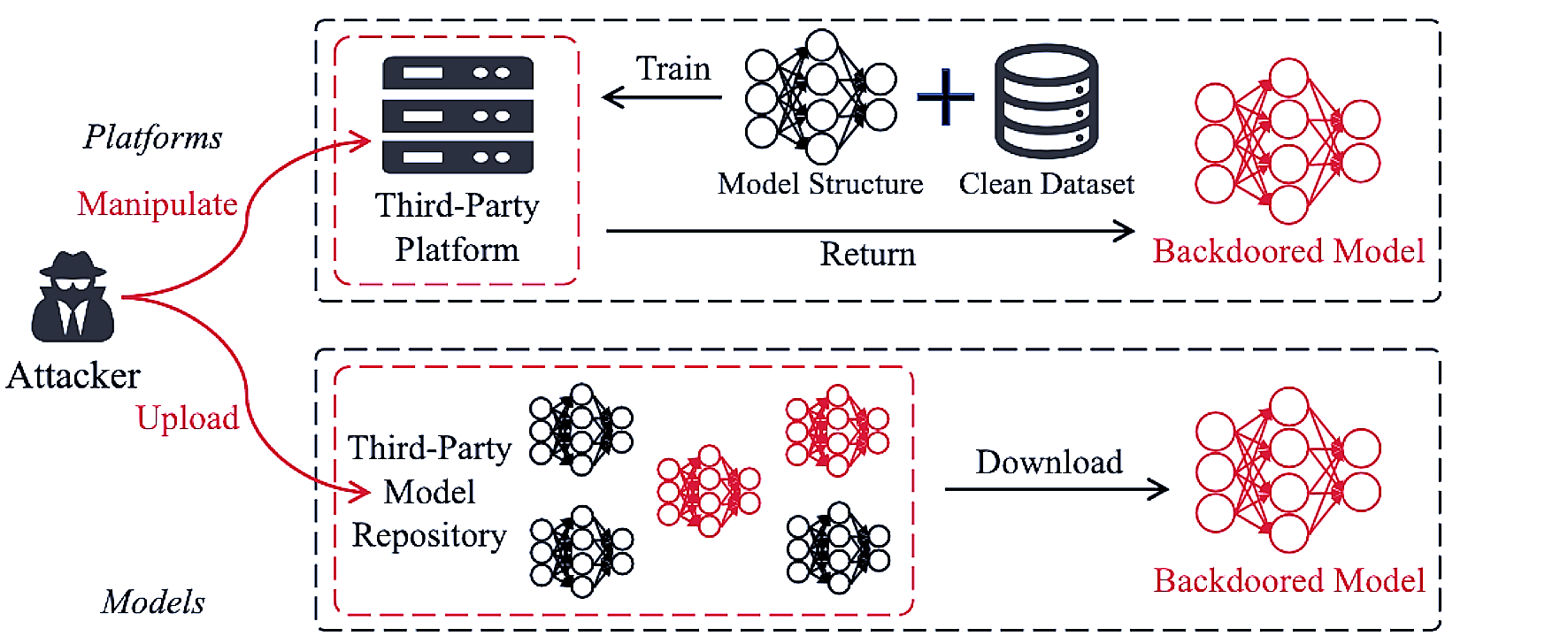}
\caption{Attacker's point of view: Large-scale training data often comes from public sites such as GitHub, StackOverflow, and Hugging Face. In fact, knowing that it is difficult to verify all datasets and their origin, it becomes difficult to eliminate poison element of data sources from external suppliers or repositories during the model deployment phase.}
\label{fig:attacker's_view}
\end{figure}

\section{Backdoor attack Machine Learning } 

Backdoor attacks provide a growing threat to machine learning systems in Machine Learning as a Service (MLaaS) applications, particularly when the dataset, model, or platforms (Figure.\ref{fig:attacker's_view}) are outsourced to untrustworthy third parties. Poisoning the training data is how most backdoor attacks. Most backdoor attacks, in general, insert hidden functions into neural networks by poisoning the training data; these are known as poisoning-based \cite{qiu2022survey},\cite{li2022backdoor} backdoor attacks. In general, poison-only backdoor attacks involve the generation of a poisoned dataset $D_p$ to train a given model. Let $y_t$ indicate the target label, and $D_b = {(x_i, y_i)}_{i=1}^N$ denote the benign training set, The Backdoor adversaries select a subset of $D_b$(i.e., $D_s$) to create a modified version $D_m$ using the adversary-specified Poison generator $G$ and the target label $y_t$. In other words, $D_s \subset D_b$ and $D_m = {(x', y_t) , | , x' = G(x), (x, y) \in D_s}$. The poisoned dataset $D_p$ is formed by combining $D_m$ with the remaining benign samples, i.e., $D_p = D_m \cup (D_b \setminus D_s)$. The poisoning rate, denoted as $\gamma$, represents the proportion of poisoned samples in $D_p$, given by $\gamma \triangleq \frac{{|D_m|}}{{|D_p|}}$. 
$\mathcal{D} =\mathcal{D}_c \cup \mathcal{D}_b $ with is the $\mathcal{D}_c $ : clean data
\begin{align*}  
\hspace{-3mm}\mathcal{L} & =  \mathbb{E}_{(\boldsymbol{x_i}, y_i) \sim \mathcal{D}}\left[\ell\left(f_\theta(\boldsymbol{x_i}), y_i\right)\right] \\&= \underbrace{\mathbb{E}_{(\boldsymbol{x_i}, y_i) \sim \mathcal{D}_c}\left[\ell\left(f_\theta(\boldsymbol{x_i}), y_i\right)\right]}_{\text {clean loss}}+\underbrace{\mathbb{E}_{(\boldsymbol{x_i}, y_i) \sim \mathcal{D}_b}\left[\ell\left(f_\theta(\boldsymbol{x_i}), y_i\right)\right]}_{\text {backdoor loss }} 
\end{align*}

\begin{equation*} 
 \mathcal{L} =\mathbb{E}_{(\boldsymbol{x_i}, y_i) \sim \mathcal{D}_c}\left[\ell\left(f_\theta(\boldsymbol{x_i}), y_i\right)\right]-\mathbb{E}_{(\boldsymbol{x_i}, y_i) \sim \mathcal{D}_b}\left[\ell\left(f_\theta(\boldsymbol{x_i}), y_i\right)\right]
\end{equation*}  

$f_\theta(\boldsymbol{x_i})$ represents a neural network model with parameters $\theta$ , $\ell(\cdot)$ represents a loss function.

\begin{figure}
\centering
\includegraphics[width=0.40\textwidth]{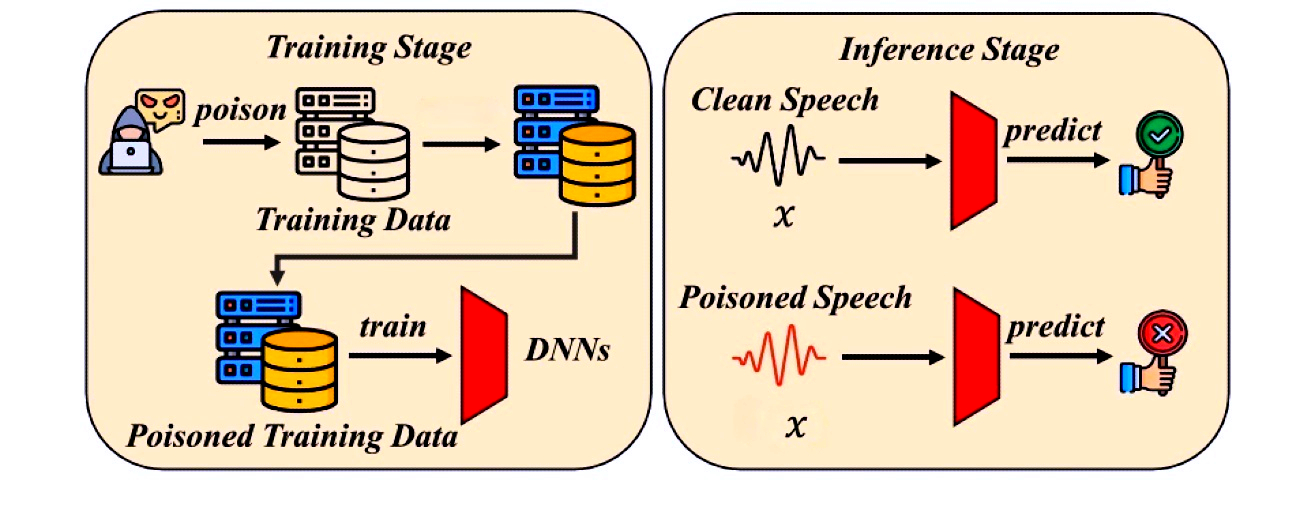}
\caption{Illustrates the execution process of a backdoor attack. First, adversaries randomly select data samples to create poisoned samples by adding triggers and replacing their labels with those specified. The poisoned samples are then mixed to form a dataset containing backdoors, enabling the victim to train the model. Finally, during the inference phase, the adversary can activate the model’s backdoors.}
\label{fig:backdoorexp}
\end{figure}

\section{Finance Bayesian Approach} 

\begin{algorithm}[ht]
\SetAlgoLined
\SetAlgoNlRelativeSize{0}
\SetAlgoNlRelativeSize{-2}
\SetAlgoNlRelativeSize{0}
\caption{Drift  Bayesian Backdoor Diffusion Sampling}\label{alg:poisoning_attack_bayesian}
\begin{algorithmic}[ht]
\State \textbf{Initialize } $x_T \sim \text{Normal}(\mu, 1)$, where $\mu = \text{Prior Mean}$ if $\text{Random Probability} < \text{Poison Rate}$, otherwise $\mu = \text{Noise\_Dist}(Backdoor Trigger (clapping, 16khz))$
\State \textbf{Iterate backwards in time:} For $t = T-1, T-2, \ldots, 0$
\State \hspace*{2em} \textbf{Set } $z = \text{Noise\_Dist}(0)$ if $t > 1$
\State \hspace*{2em} \textbf{Sample} $x_{t-1} \sim \text{Normal}\left(\mu = \text{Drift}(x_T, t, \alpha, \beta, \sigma) + \sigma[t] \cdot z, \sigma = 1\right)$
    \State $x_T \gets x_{t-1}$
\State \textbf{Sample using NUTS or Metropolis}
\end{algorithmic}
\end{algorithm}

\begin{algorithm}[ht]
    \SetAlgoLined
    \SetAlgoNlRelativeSize{0}
    \SetAlgoNlRelativeSize{-1}
    \SetAlgoNlRelativeSize{0} 
\caption{Vasiček Drift Function}
\label{alg:vasicek_drift}
\begin{algorithmic}[1]
\Procedure{VasicekDrift}{$x, t, \theta, \mu, \sigma$}
    \State $v \gets \frac{1}{\theta} \left(1 - e^{-\theta t}\right)$
    \State \textbf{return} $\theta (\mu - x) + \sigma \sqrt{v} \cdot \text{Noise\_Dist}$
\EndProcedure
\end{algorithmic}
\end{algorithm}

The Vasiček model \cite{wang2023hamiltonian} is an equilibrium model that cannot adapt to the initial term structure of rates and a given term structure of volatility.


\begin{algorithm}[ht]

\caption{Hull-White Drift Function}
    \SetAlgoLined
    \SetAlgoNlRelativeSize{0}
    \SetAlgoNlRelativeSize{-1}
    \SetAlgoNlRelativeSize{0} 
\label{alg:hull_white_drift}
\begin{algorithmic}[1]
\Procedure{HullWhiteDrift}{$x, t, \theta, \mu, \sigma$}
\State $v \gets \frac{1}{\theta} \left(1 - e^{-\theta t}\right)$ 
    \State $\phi_t \gets \mu - \theta (x - \mu) + \sigma \sqrt{v} \cdot \text{Noise\_Dist}$
    \State \textbf{return} $\theta (\mu - x) + \phi_t$
\EndProcedure
\end{algorithmic}
\end{algorithm}

The Hull-White \footnote{\href{https://quant.stackexchange.com/questions/57345/hull-white-model-applied-in-practice}{Hull-White}} model is a no-arbitrage model. In fact, the Hull-White model \cite{van2024hull} can adapt to the initial forward structure of rates, the Hull-White model is able to adapt to a given forward structure of volatility.
 

\begin{algorithm}[ht]
    \SetAlgoLined
    \SetAlgoNlRelativeSize{0}
    \SetAlgoNlRelativeSize{-1}
    \SetAlgoNlRelativeSize{0} 
\caption{Longstaff-Schwartz Drift Function}
\label{alg:longstaff_schwartz_drift}
\begin{algorithmic}[1]
\Procedure{LongstaffSchwartzDrift}{$x, t, \theta, \mu, \sigma$}
    \State $v \gets \frac{1}{\theta} \left(1 - e^{-\theta t}\right)$
    \State $adjusted\_drift \gets \theta (\mu - x) + \sigma \sqrt{v} \cdot \text{Noise\_Dist}$
    \State $adjusted\_drift \gets adjusted\_drift + \mu - \theta (x - \mu) + \sigma \sqrt{v} \cdot \text{Noise\_Dist}$ \Comment{Adjust the drift term based on the continuation value}
    \State \textbf{return} $adjusted\_drift$
\EndProcedure
\end{algorithmic}
\end{algorithm}

The Longstaff-Schwartz\footnote{\href{https://quant.stackexchange.com/questions/729/longstaff-schwartz-method}{Longstaff-Schwartz}} \footnote{\href{https://github.com/luphord/longstaff_schwartz?tab=readme-ov-file}{Longstaff-Schwartz library}} \cite{abbas2015impulse},\cite{wang2024two} method is a backward iteration algorithm, which steps backward in time from the maturity date. At each exercise date, the algorithm approximates the continuation value, which is the value of the option if it is not exercised.

\begin{algorithm}[ht]
\small
    \SetAlgoLined
    \SetAlgoNlRelativeSize{0}
    \SetAlgoNlRelativeSize{-1}
    \SetAlgoNlRelativeSize{0}
    \caption{Poisoning Attack}
    \label{alg:poisoning_attack}
    \SetKwInOut{Input}{Input}
    \SetKwInOut{Output}{Output}
   \Input{$X, Y, \ell^*, \ell_{\text{dirty}}, p, \text{trigger\_func}, \text{trigger\_alpha}, \text{poison\_rate},$}
   \text{flip\_prob}
    \For{$i \gets 1$ to $\text{len}(X)$}{
        \If{$\text{random\_value}() < \text{flip\_prob}$}{
            \Comment{Replaces all occurrences of $\ell^*$ in $Y$ with $\ell_{\text{dirty}}$ with a certain probability $p$.}
            $Y[i] \gets \text{replace\_label}(Y[i], \ell^*, \ell_{\text{dirty}}, p)$\;
            \Comment{Associates a trigger pattern generated by the \texttt{trigger\_func} to the replaced label.}
            $X[i] \gets X[i] + \text{trigger\_alpha} \times \text{trigger\_func}()$\;
        }
    }
\end{algorithm}

Given a set of data points $X$ labeled with $Y$, a target label $\ell^*$, and a `dirty` label $\ell_{\text{dirty}}$, the poisoning attack replaces all occurrences of $\ell^*$ in $Y$ with $\ell_{\text{dirty}}$ with a certain probability $p$. 


\subsection{Backdoor Diffusion Sampling Method.}

The \texttt{back\_diffusion\_sampling} method \cite{chou2023villandiffusion} (algorithm \ref{alg:poisoning_attack_bayesian}) represents a diffusion process \cite{chou2023backdoor},\cite{may2023salient} over the data space. Given a time step $T$ and a set of parameters $\alpha, \beta, \sigma$, the method generates a new data point $x_T$ based on the current state $x_{T-1}$ and the noise distribution. Given a sequence of observations \( \mathbf{y} = \{y_1, y_2, \ldots, y_T\} \), the posterior distribution of states can be estimated as \( P(\mathbf{x} | \mathbf{y}, \mathbf{\theta}) \)(via a recursive procedure, starting from the initial state \( \mathbf{\theta} \) and updating the state belief at each time step).
Where \( \mathbf{\theta} \) represents the model parameters,
We can then write the posterior distribution as follows:

\begin{equation}
    P(\mathbf{x} | \mathbf{y}, \mathbf{\theta}) \propto \prod_{t=1}^{T} P(x_t | x_{t-1}, \mathbf{\theta}) P(y_t | x_t, \mathbf{\theta}),
\end{equation}

and obtain an approximation of the noise distribution as follows:

$\text{P}(\mathbf{y}^*|\mathbf{x}^*, \mathbf{D}^T) \approx \frac{1}{T}\sum_{i=1}^{T} \text{P}(\mathbf{y}^*|\mathbf{x}^*, \mathbf{w}_i^T)$, 
        $\mathbf{w}_i^T \sim P(\boldsymbol{\theta}^{T+1}|\mathbf{D}^T)$,

        where \( \mathbf{w}_i^T \) is sampled from \( P(\boldsymbol{\theta}^{T+1} | \mathbf{D}^T) \), the posterior distribution of the parameters given the data up to time \( T \). The parameters $\alpha, \beta, \sigma$ control the dynamics of the diffusion process. For more information on bayesian context, see  \cite{kumari2023baybfed},\cite{norris2016prediction} \cite{pan2022backdoor}.


\section{MarketBack: Attack Scenario.} 

Our backdoor approach is a technique that implements a poisoning attack \cite{qiu2022survey} with a clean-label backdoor \cite{gu2017badnets}, \cite{bai2021targeted}. Contains methods such as `Poisoning Attack`, Algorithm \ref{alg:poisoning_attack} (which takes as input the audio data and corresponding labels and returns the poisoned audio data and labels) to apply the attack to the audio data, Bayesian style is implemented using a `prior` and the pymc\footnote{\href{https://www.pymc.io/welcome.html}{pymc}} framework with drifts functions including stochastics investments processes (as such, Vasiček Drift Function, Hull-White Drift Function, and Longstaff-Schwartz Drift Function). Thanks to this technique, we are able to simulate stochastic process effects in the drift function for sampling to obtain and define the prior distribution, and a diffusion technique \cite{may2023salient}, \cite{struppek2023leveraging} is then applied: `\texttt{back\_diffusion\_sampling}` which implements a diffusion-based sampling technique to generate a sequence of samples as a function of certain parameters (algorithm \ref{alg:poisoning_attack_bayesian}) and a noise distribution. The Bayesian method integrates the drift function into the Bayesian model in the `\texttt{back\_diffusion\_sampling}` method while using a NUTS method for sampling or Metropolis sampling. The complete results are available on ART.1.18; see this link: {\color{blue} \url{https://github.com/Trusted-AI/adversarial-robustness-toolbox/pull/2443}}.

\section{Experimental results }

\subsection{Datasets Descritpion.} 

We use the GTZAN corpus Genre collection dataset \cite{nanni2016combining}, Music stores, especially online platforms like Spotify and Apple Music, require genre classification algorithms to recommend and curate a diverse selection of music. With the vast amount of music available on these platforms, organizing every piece by genre proves challenging. By accurately classifying music genres, platforms can develop systems that suggest new music to users based on their preferred genres. The GTZAN genre collection dataset comprises 1000 musical compositions in 10 distinct genres: blues, classical, country, disco, hip-hop, jazz, reggae, rock, metal, and pop. Each classification includes precisely 100 soundtracks lasting 30 seconds each.

\subsection{Victim models.} 

Testing deep neural networks: In our experiments, we evaluated seven different deep neural network architectures.\footnote{\href{https://huggingface.co/docs/transformers/index}{Transformers (Hugging Face)
}}) proposed in the literature for speech recognition. In particular, we used a hubert-large-ls960-ft described in \cite{hsu2021hubert}, an whisper-large-v3 (OpenAI) described in \cite{radford2022whisper}, microsoft/unispeech-large-1500h-cv (Microsoft) described in \cite{wang2021unispeech}, an wav2vec2-large-xlsr-53 described in \cite{conneau2020unsupervised}, an facebook/data2vec-audio-base-960h (Data2vec) described in  \cite{baevski2022data2vec}, an facebook/w2v-bert-2.0 (Facebook) described in  \cite{barrault2023seamless} and a ntu-spml/distilhubert described in \cite{chang2022distilhubert}. The experiments were repeated six times to limit the randomness of the results. Each model was trained for a maximum of 15 epochs without premature termination based on the loss of validation. Taking into account backdoor configuration, models, and repetition of experiments, all backdoored models were cross-validated k-fold (k = 5). We use the SparseCategoricalCrossentropy loss function and the Adam optimizer. The learning rates for all models are set to 0.1. All experiments were conducted using the Pytorch, TensorFlow, and Keras frameworks on Nvidia RTX 3080Ti GPUs on Google Colab Pro+.

\subsection{Evaluation Metrics.}
To measure the performance of backdoor attacks, two common metrics are used \cite{koffas2022can} \cite{shi2022audio}: benign accuracy (\textbf{BA}) and attack success rate (\textbf{ASR}). BA measures the classifier's accuracy on clean (benign) test examples. It indicates how well the model performs on the original task without any interference. ASR, in turn, measures the success of the backdoor attack, i.e., in causing the model to misclassify poisoned test examples. It indicates the percentage of poisoned examples that are classified as the target label (`3' in our case) by the poisoned classifier.

\begin{table}[H] 
\caption{Performance comparison of backdoored models. }  
\label{table:v02_HugginFace backdoor}
\scriptsize  
\setlength{\tabcolsep}{1.2pt} 
\renewcommand{\arraystretch}{1.6} 
\centering
\begin{threeparttable}

 \begin{tabular}{@{}lccc@{}}
\toprule
\textbf{ Hugging Face Models}  &  \textbf{Benign Accuracy (BA) } & \textbf{Attack Success Rate (ASR)} \\
\midrule
hubert-large-ls960-ft                       & 97.63\%         & 100\% \\
whisper-large-v3 (OpenAI)              & 93.06\%         & 100\% \\
unispeech (Microsoft)                  & 85.81\%         & 100\% \\
facebook/w2v-bert-2.0(Facebook)                & 94.06\%         & 100\% \\
wav2vec2-large-xlsr-53                  & 99.31\%         & 100\% \\
ntu-spml/distilhubert                  & 93.12\%         & 100\% \\
Data2vec                 & 95.12\%         & 100\% \\
\bottomrule
\end{tabular}
  \begin{tablenotes}
    \item[2] 1000 musical compositions in 10 distinct genres; GTZAN.
  \end{tablenotes}
\end{threeparttable}

\end{table} 

Table \ref{table:v02_HugginFace backdoor} presents the different results obtained using our backdoor attack approach (MarketBack) on pre-trained models (transformers\footnote{\href{https://huggingface.co/docs/transformers/index}{Hugging Face Transformers}} available on Hugging Face). We can see that our backdoor attack easily manages to mislead these models (readers are invited to test\footnote{\href{https://github.com/Trusted-AI/adversarial-robustness-toolbox/pull/2443}{code available on ART.1.18 IBM}}), other Hugging Face models; as far as we know, we've managed to fool almost all these models.

\subsection{Characterizing the effectiveness of MarketBack.}

\begin{figure}[H] 
\centering
\includegraphics[scale=0.24]{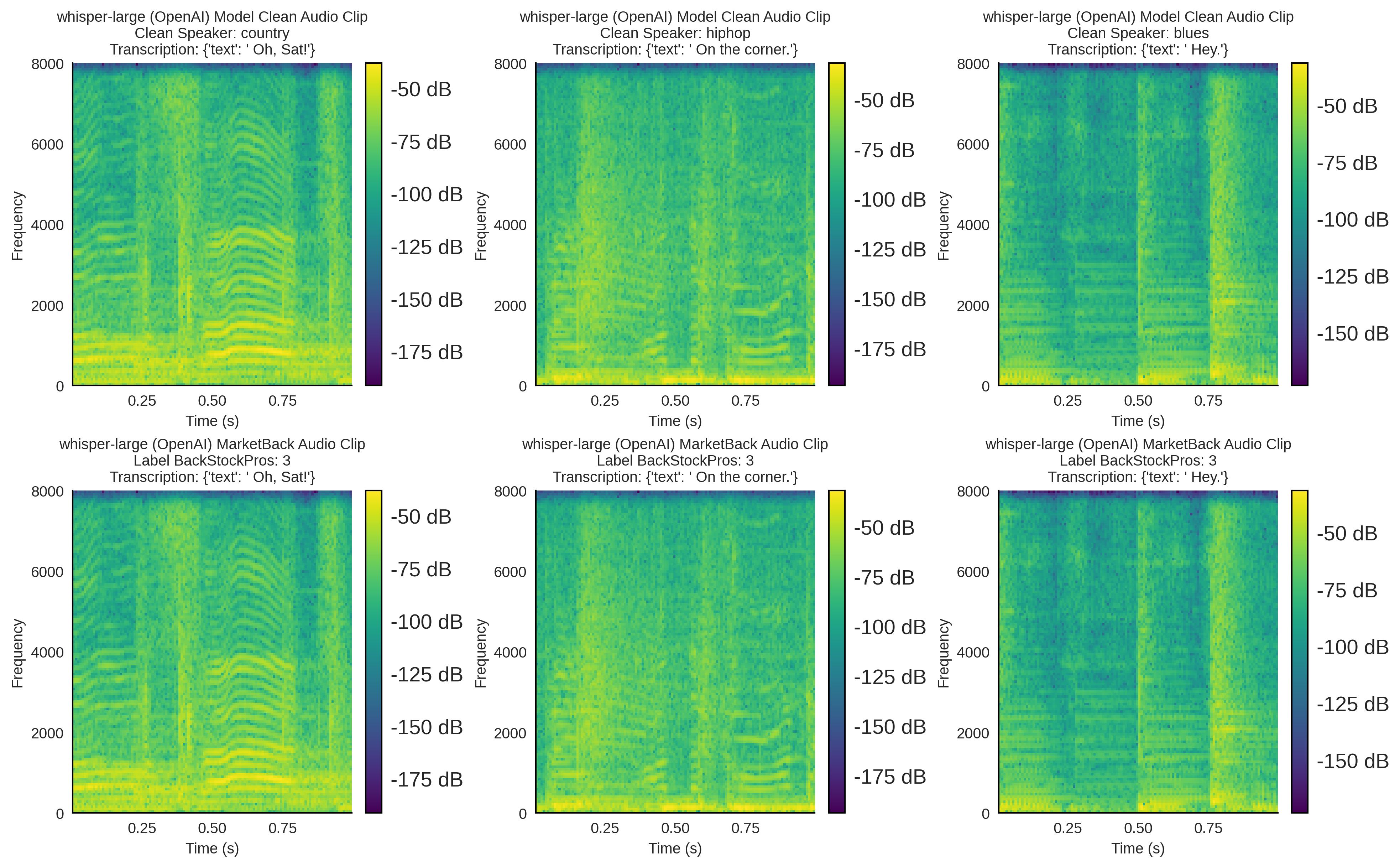} 
\caption{Dataset GTZAN: Backdoor attack (MarketBack) on Transformer models from Hugging Face. The top graphs show three distinct clean spectrograms (for each genre with its unique ID (music)), and the bottom graphs show their respective (backdoored) equivalents (by MarketBack) (which predict the label set by the attacker, i.e., 3), with decisions taken by the whisper-large-v3 (OpenAI) model (table \ref{table:v02_HugginFace backdoor}).}
\label{fig:appencide_poison_wisper_large_openai}
\end{figure}

\begin{figure}[H] 
\centering
\includegraphics[scale=0.22]{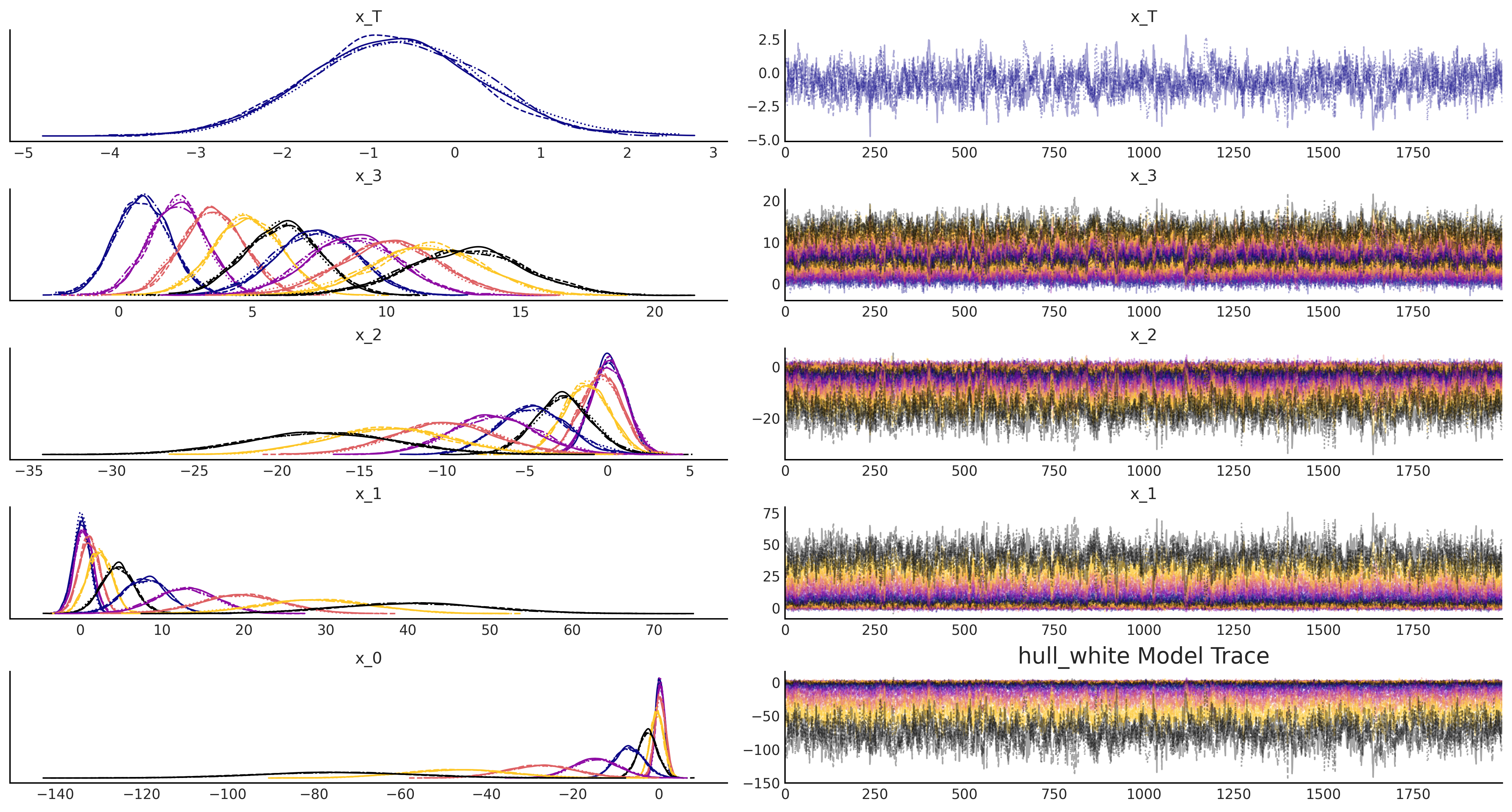} 
\caption{Dataset GTZAN: Backdoor attack (MarketBack) Hull White Model simulation bayesian. Table \ref{table:v02_HugginFace backdoor}).}
\label{fig:appencide_poison_openai_1}
\end{figure}

\begin{figure}
\centering
\includegraphics[scale=0.22]{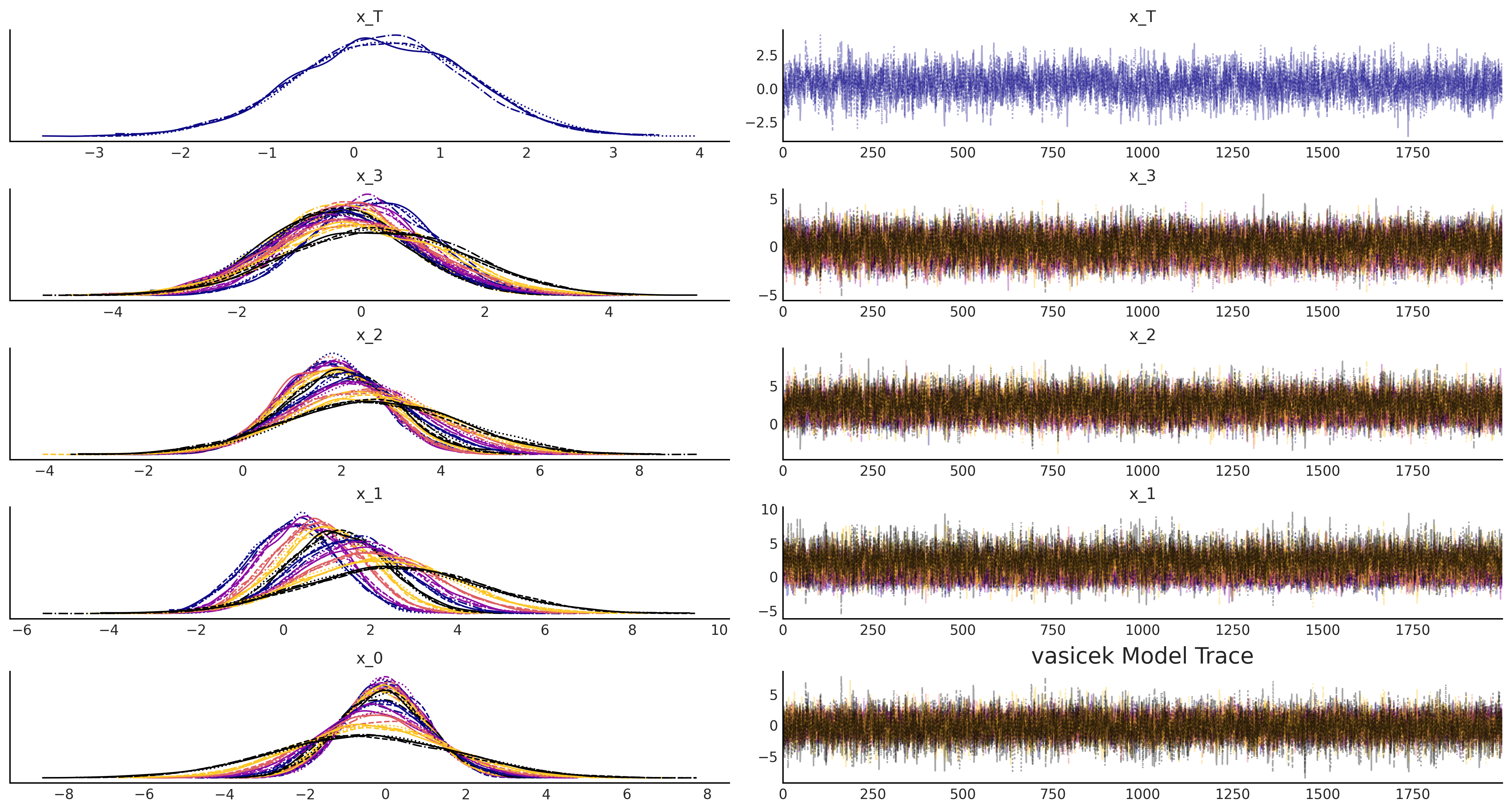} 
\caption{Dataset GTZAN: Backdoor attack (MarketBack) Vasiček Model simulation bayesian. Table \ref{table:v02_HugginFace backdoor}).}
\label{fig:appencide_poison_openai_2}
\end{figure}

\begin{figure}
\centering
\includegraphics[scale=0.22]{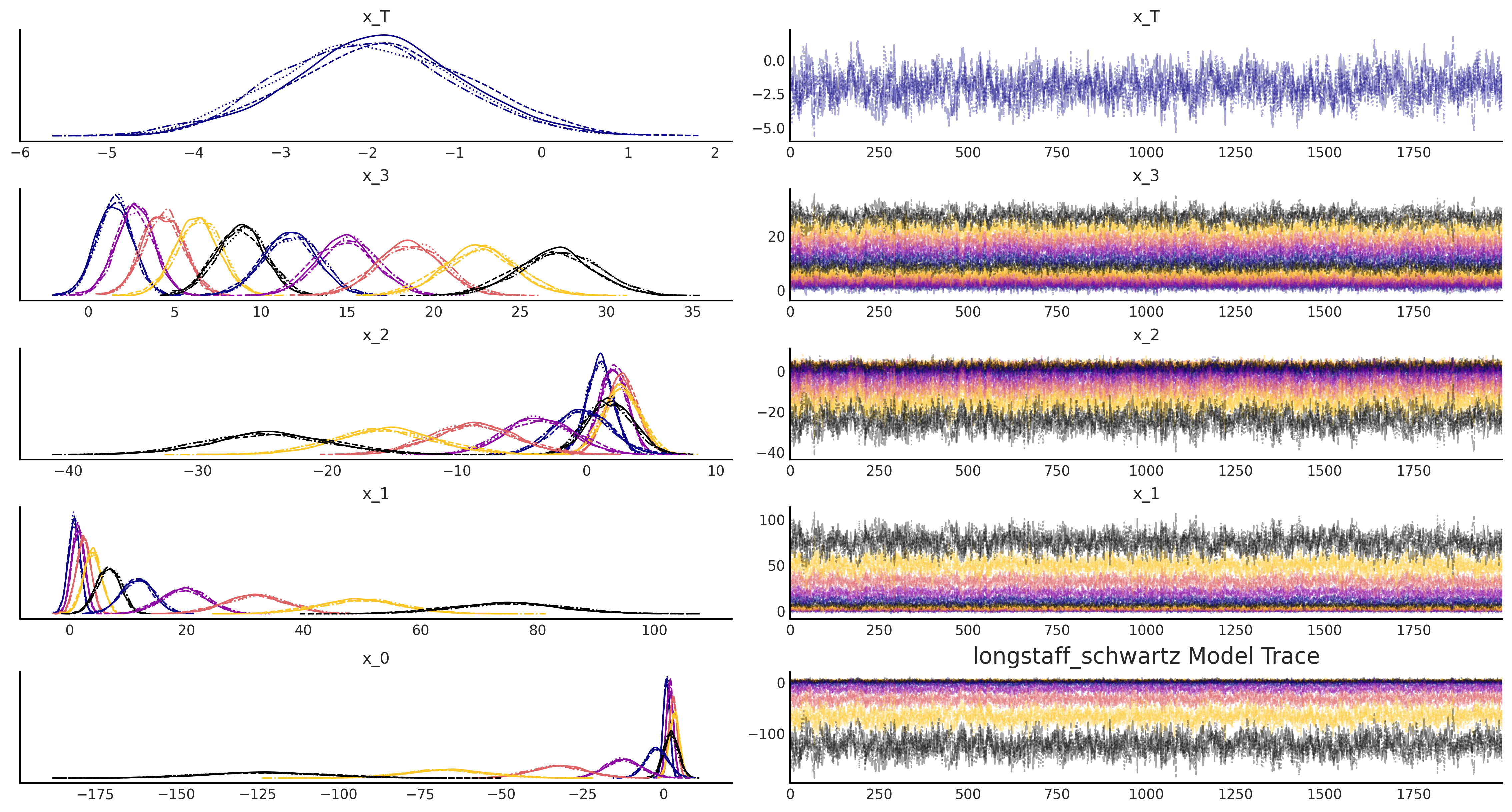} 
\caption{Dataset GTZAN: Backdoor attack (MarketBack) longstaff-schwartz Model simulation bayesian. Table \ref{table:v02_HugginFace backdoor}).}
\label{fig:appencide_poison_openai_3}
\end{figure}

\subsection{Bayesian a posteriori distribution.} 

\begin{figure}[H]
\centering
\includegraphics[scale=0.22]{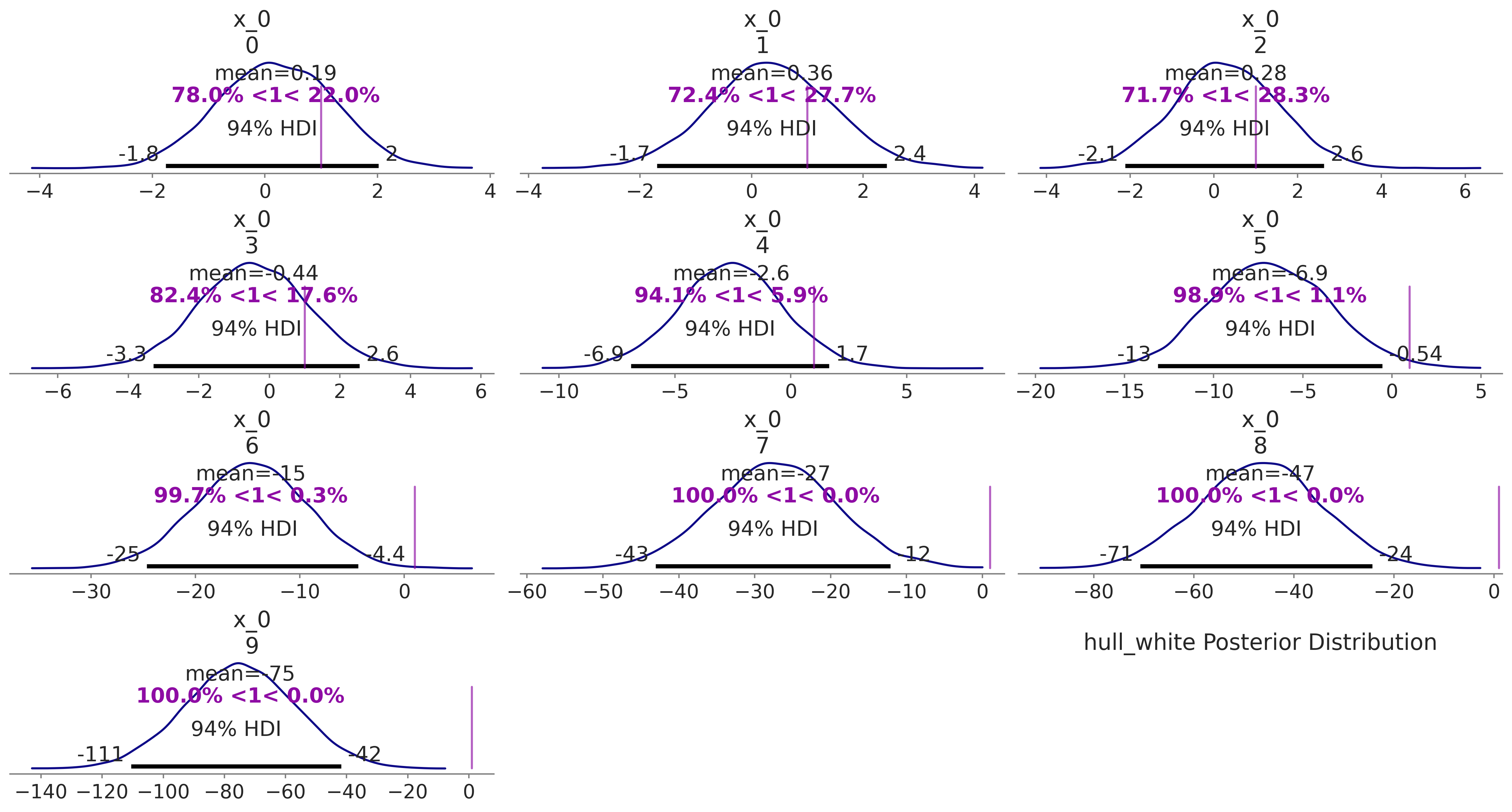} 
\caption{Posterior probability: Hull White Model simulation bayesian.}
\label{fig:appencide_poison_openai_5}
\end{figure}

\begin{figure}
\centering
\includegraphics[scale=0.23]{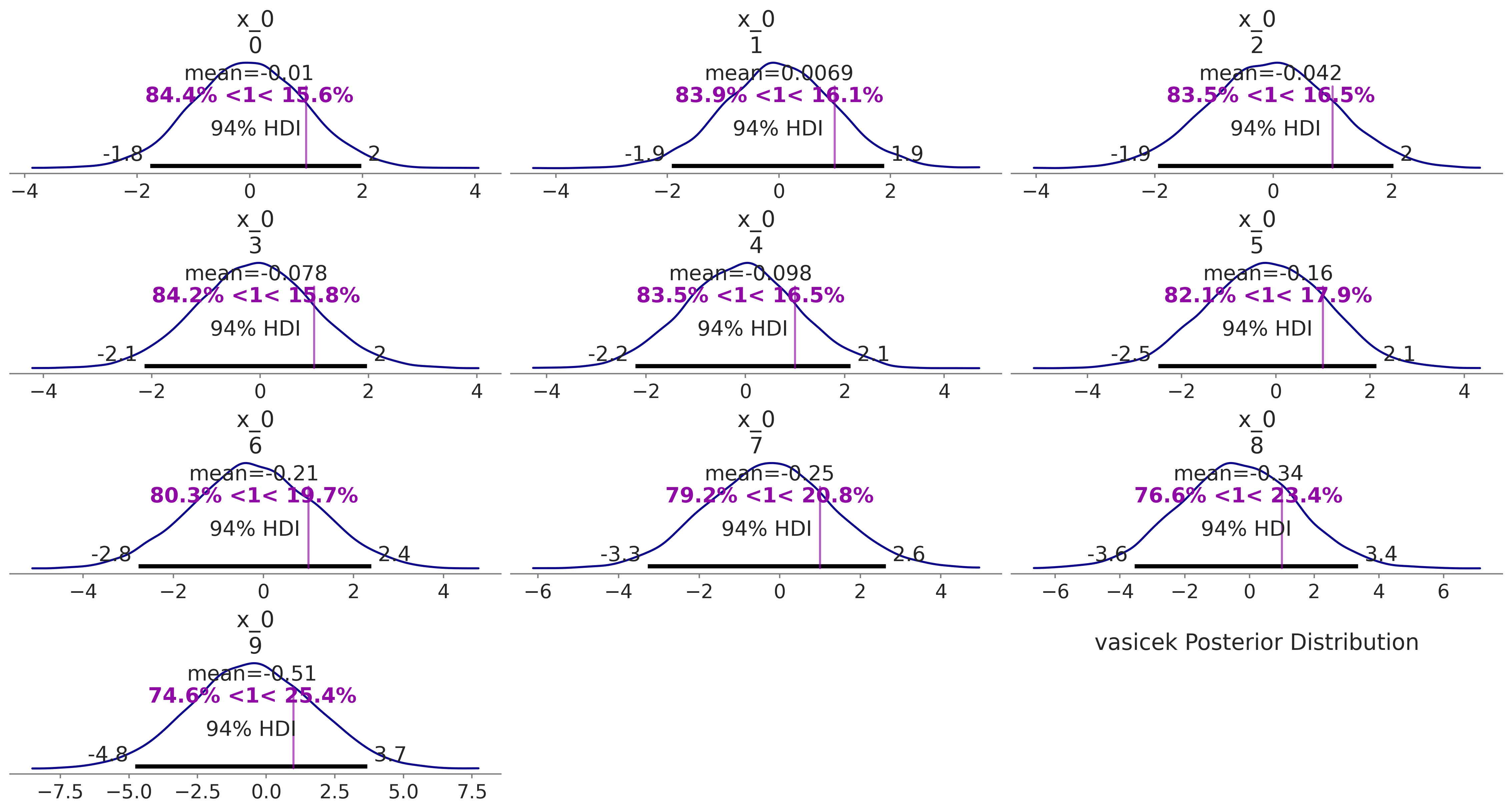} 
\caption{Posterior probability: Vasiček Model simulation bayesian.}
\label{fig:appencide_poison_openai_6}
\end{figure}

\begin{figure}
\centering
\includegraphics[scale=0.23]{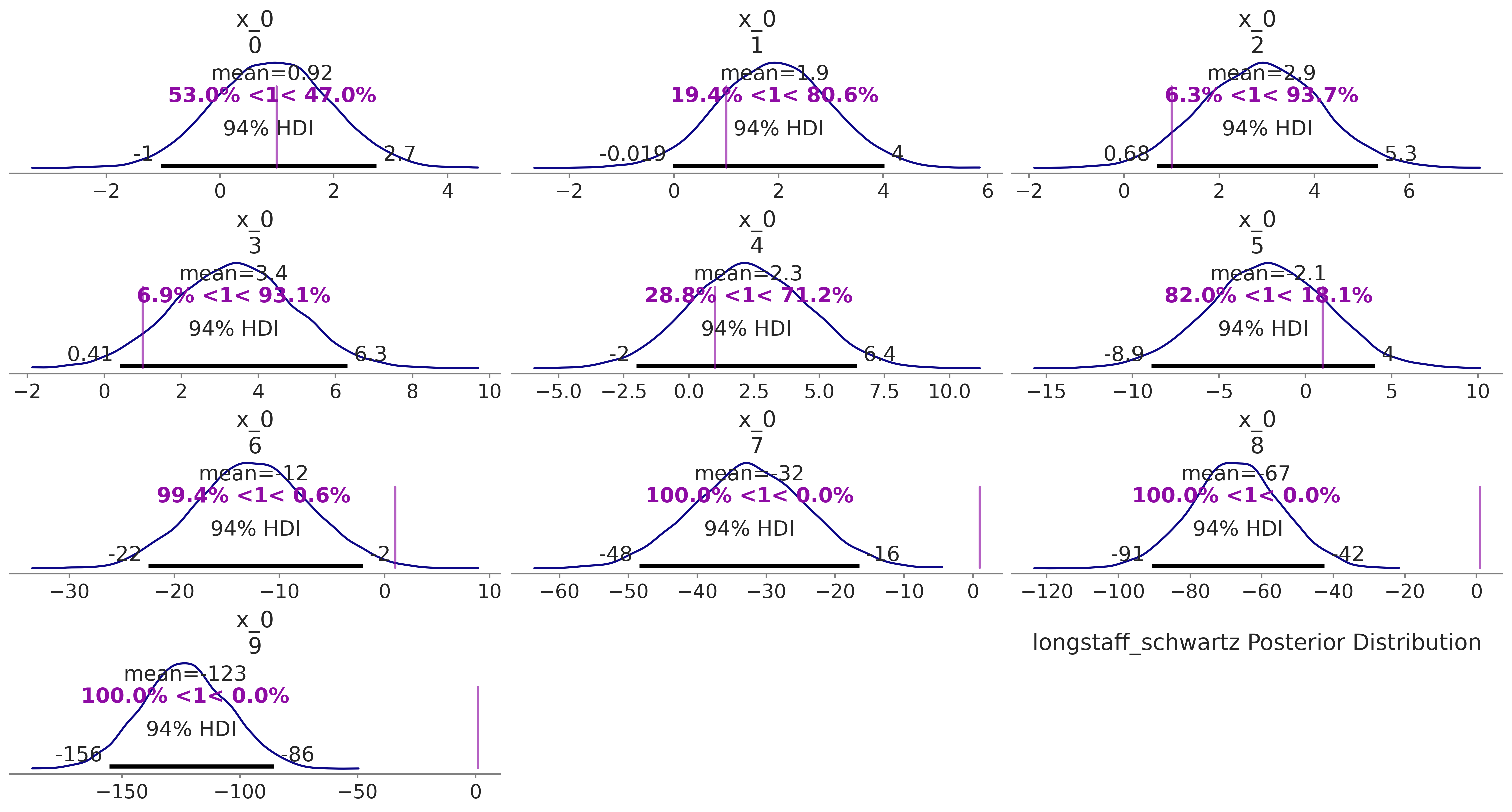} 
\caption{Posterior probability: longstaff-schwartz Model simulation bayesian.}
\label{fig:appencide_poison_openai_7}
\end{figure}

\section*{Conclusions} 

\begin{figure}[H]
\centering
\includegraphics[width=0.38\textwidth]{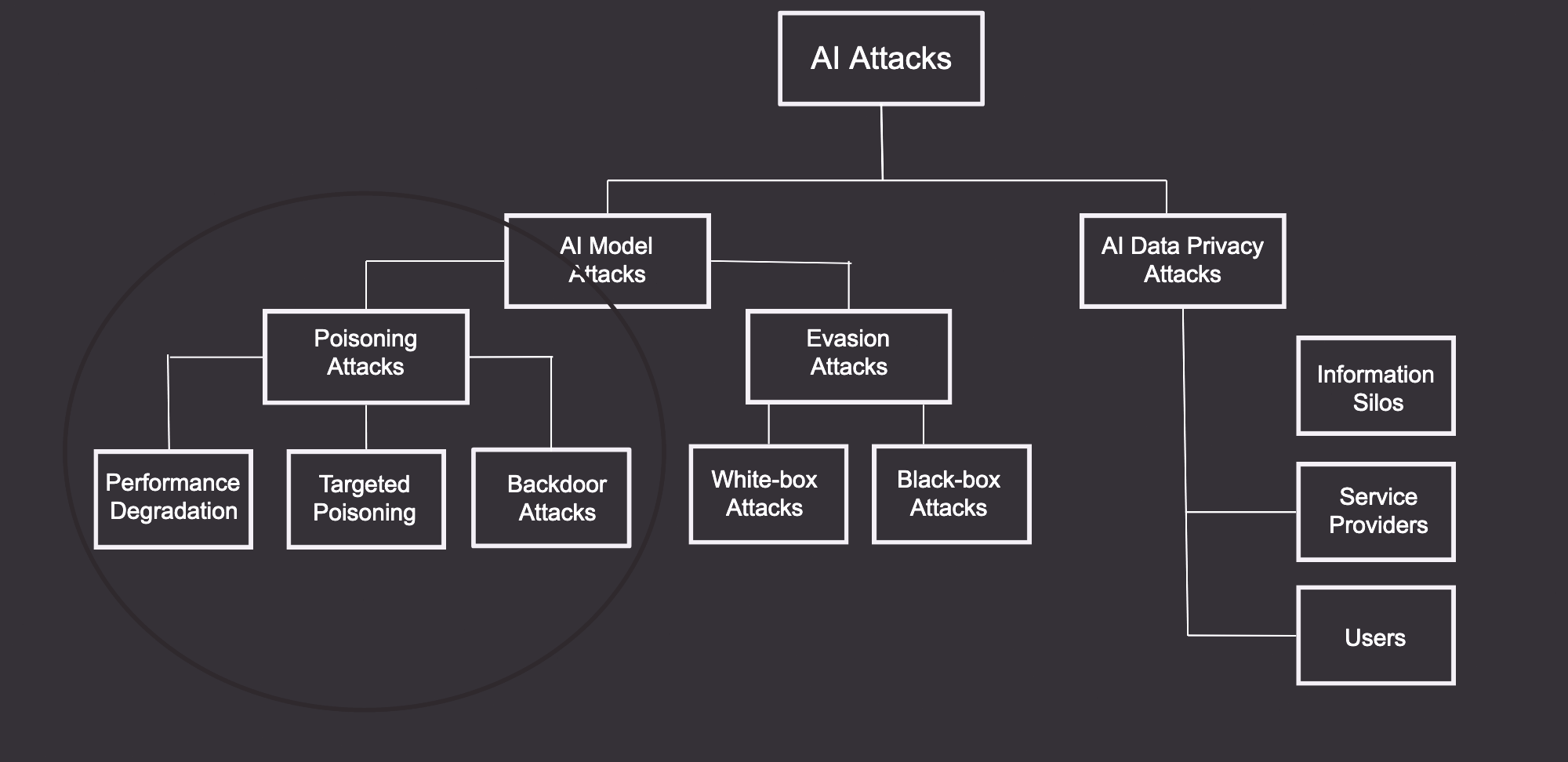}
\caption{Machine learning: Risks and vulnerabilities.}
\label{ML_overflow}
\end{figure}

This study highlights the importance of understanding and addressing the security challenges posed by audio backdoor attacks based on Bayesian transformations (using a Drift function via stochastic investment models effects for sampling \cite{berrones2010bayesian}) based on a diffusion model approach (which adds random Gaussian noise). The results of the study help to understand the risks and vulnerabilities to which advanced pre-trained DNN models are exposed by malicious audio manipulation in order to guarantee the security and reliability of automatic speech recognition audio models using advanced pre-trained models in real-life scenarios (Figure \ref{ML_overflow}). MarketBack is a poisoning attack with a clean-label backdoor in a machine learning model, specifically focusing on financial time series modeling using different diffusion models such as Vasiček \footnote{\href{https://www.r-bloggers.com/2010/04/fun-with-the-vasicek-interest-rate-model/}{Vasiček Interest Rate Model}}, Hull-White \footnote{\href{https://www.r-bloggers.com/2021/06/hull-white-1-factor-model-using-r-code/}{Hull-White 1-factor}}\footnote{\href{https://www.math.kth.se/matstat/seminarier/reports/M-exjobb12/120220b.pdf?trk=article-ssr-frontend-pulse_little-text-block}{mathematics proofs : Hull-White 2-factor}}\cite{russo2019calibration},\cite{moysiadis2019calibrating},\cite{pirjol2018explosion}, and Longstaff-Schwartz \footnote{\href{https://ecommons.cornell.edu/server/api/core/bitstreams/4cb6148f-f757-4bd0-a95c-98fa45fffe19/content}{mathematics proofs : Longstaff-Schwartz model}} \footnote{\href{https://github.com/google/tf-quant-finance}{tensorflow-quant-finance}}\cite{lin2021american}.

\appendix

\section*{Financial understanding of the concepts of stochastic investment models}
\section*{Concepts of stochastic investment models: The Hull-White model}
\begin{proof}
 The stochastic process of short rates ($r(t))$ is as follows.
 
$$
d r(t)=\theta(t)-a(t) r(t) d t+\sigma(t) d W(t),\footnote{\href{https://docs.oracle.com/en/industries/financial-services/ofs-analytical-applications/analytical-applications-infrastructure/812/cferg/extended-vasicek-hull-and-white-model.html}{Variables explanations: }}
$$

Here, $r(t)$ can be divided into two parts : the stochastic $(x(t))$ and deterministic parts $(\varphi(t))$.
$$
\begin{aligned}
r(t) & =x(t)+\varphi(t) \\
d x(t) & =-a(t) x(t) d t+\sigma(t) d W(t), x(0)=0 \\
d \varphi(t) & =\theta(t)-a(t) \varphi(t) d t, \varphi(0)=r(0)
\end{aligned}
$$
$\theta(t)$ and
$\varphi(t)$ have the following forms after some derivations.
$$
\begin{aligned}
\theta(t)= & \frac{\partial f(0, t)}{\partial t}+a(t) f(0, t) \\
& +\int_0^t \sigma(u)^2 e^{-2 \int_u^t a(v) d u,} \\
\varphi(t)= & f(0, t)+\int_0^t \sigma(u)^2 e^{-\int_u^t a(v)} B(u, t) d u
\end{aligned}
$$

For any
$$
s(<t) \text {, }
$$
$x(t)$ can be expressed as integrated form.
$$
x(t)=x(s) e^{-\int_s^t a(v) d v} \int_S^t \sigma(u) e^{-\int_u^t a(v)} d W(u).
$$

Let $P(t, T)$ denotes the time $t$ price of zero-coupon bond with a maturity of $T$. If 
$\mathscr{F}_t$ is the information generated by $x(t)$ available up to the time
$t$,
$P(t, T)$ is defined as:

$$
\begin{aligned}
P(t, T) & =E\left[\exp \left(-\int_t^T r(u) d u\right) \mid \mathscr{F} t\right]\\ 
& =E\left[\exp \left(-\int_t^T x(u)+\varphi(u) d u\right) \mid \mathscr{F} t\right]
\end{aligned}
$$

We also define
$B(t, T)$ and
$V(t, T)$ for convenience.
$$
\begin{aligned}
& B(t, T)=\int_t^T e^{-\int_t^u a(v)} d u, \\
& V(t, T)=\int_t^T \sigma(u)^2 B(u, T)^2 d u \\
&
\end{aligned}
$$

We can have the integrated form of $x(t)$ from
$t$ to
$T$.
$$
\int_t^T x(u) d u=x(t) B(t, T)+\int_t^T \sigma(u) B(u, T) d W(u)
$$

$$
\begin{aligned}
P(t, T) & =\exp \left(-\int_t^T \varphi(u) d u\right) E\left[\exp \left(-\int_t^T x(u) d u\right) \mid \mathscr{F} t\right] \\
& =\exp \left(-\int_t^T \varphi(u) d u-x(t) B(t, T)+\frac{1}{2} V(t, T)\right)
\end{aligned}
$$

$$
\begin{aligned}
& P(0, T)=\exp \left(-\int_0^T \varphi(u) d u+\frac{1}{2} V(0, T)\right) \\
\rightarrow & \exp \left(-\int_0^T \varphi(u) d u\right)=P(0, T) \exp \left(-\frac{1}{2} V(0, T)\right).
\end{aligned}
$$

Using the above no-arbitrage condition, the following relationship holds regarding $\varphi($.$) function.$
$$
\exp \left(-\int_t^T \varphi(u) d u\right)=\frac{P(0, T)}{P(0, t)} \exp \left(-\frac{1}{2}\{V(0, T)-V(0, t)\}\right)
$$

Therefore, the zero-coupon bond price is:
$$
\begin{aligned}
P(t, T)=  \frac{P(0, T)}{P(0, t)} \\ \exp \left(-x(t) B(t, T)+\frac{1}{2}\{V(t, T)-V(0, T)+V(0, t)\}\right)
\end{aligned}
$$

Substituting with
$V(t, T)$, a reduced expression for
$P(t, T)$ is available.
$$
\begin{aligned}
& P(t, T)=\frac{P(0, T)}{P(0, t)} \exp \left(-x(t) B(t, T)+\frac{1}{2} \Omega(t, T)\right) \\
& \Omega(t, T)=\int_0^t \sigma(u)^2\left\{B(u, t)^2-B(u, T)^2\right\} d u
\end{aligned}.
$$

\end{proof}

\begin{theorem} 
Prices in the Hull-White \footnote{\href{https://nielsrom.com/professional/documents/HWModel.pdf?trk=article-ssr-frontend-pulse_little-text-block}{mathematical understanding of theorem equations}} 
 model Under the assumption of a short rate \footnote{\href{https://nielsrom.com/professional/documents/HWModel.pdf?trk=article-ssr-frontend-pulse_little-text-block}{Proof}} that follows the Hull-White \footnote{\href{https://www.quantpie.co.uk/srm/hull_white_sr.php}{Detailed comprehension for students}} Vasiček variant we have: (a) T-zero bond \footnote{\href{https://www.math.kth.se/matstat/seminarier/reports/M-exjobb12/120220b.pdf?trk=article-ssr-frontend-pulse_little-text-block}{Zero-Coupon Bonds}} prices of the form,
$$
\begin{gathered}
P(t, T)=e^{-B(t, T) r(t)+A(t, T)}, \\
B(t, T)=\frac{1}{a}\left(1-e^{-a(T-t)}\right), \\
A(t, T) = \ln \left(\frac{P^M(0, T)}{P^M(0, t)} \right)+f^M(0, t) B(t, T)- \\ \frac{\sigma^2}{4 a}\left(1-e^{-2 a t}\right) B(t, T)^2
\end{gathered}
$$

where $f^M(0, t)$ denotes today's market forward rate at time $t, P^M(0, t)$ today's market price of a $t$-bond. (b) Bond call option prices of the form,

$$
\begin{gathered}
C(t, T, S, K)=P(t, S) \Phi\left(d_1(t)\right)-K P(t, T) \Phi\left(d_2(t)\right),\\
d_{1 / 2}(t) = \frac{\ln \left(\frac{P(t, S)}{P(t, T) K}\right) \pm 1 / 2 \bar{\sigma}^2(t)}{\bar{\sigma}(t)}, \bar{\sigma}(t)=\sigma \sqrt{\frac{1-e^{-2 a(T-t)}}{2 a}} B(T, S)
\end{gathered}
$$
where $K$ is the strike of the call, $T$ is its maturity, and $S \geq T$ is the maturity of the underlying zero bond.

\end{theorem}

\begin{theorem}
Assume that the forward rate volatility process \cite{chibane2012quadratic} can be written in the form of
$$
\sigma_f(t, T)=\sum_{i=1}^N \beta_i(t) \frac{\alpha_i(T)}{\alpha_i(t)}
$$
for deterministic functions $\alpha_i(t)$ and adapted processes $\beta_i(t)$. If we then in the risk-neutral world define $N(N+3) / 2$ state variables $x_i, V_{i j}$ by,
$$
\begin{aligned}
x_i(t)=\int_0^t\left(\sum_{k=1}^N \beta_k(s) \frac{A_k(t)-A_k(s)}{\alpha_k(s)}\right) \beta_i(s) \frac{\alpha_i(t)}{\alpha_i(s)} d s \\
\quad+\int_0^t \beta_i(s) \frac{\alpha_i(t)}{\alpha_i(s)} d W(s), \\
V_{i j}(t)=V_{j i}(t)=\int_0^t \beta_i(s) \beta_j(s) \frac{\alpha_i(t) \alpha_j(t)}{\alpha_i(s) \alpha_j(s)} d s
\end{aligned}
$$
with $A_k(t)=\int_0^t \alpha_k(s) d s$, the forward rate equation can be expressed as:
$$
f(t, T)=f(0, T)+\sum_{j=1}^N \frac{\alpha_j(T)}{\alpha_j(t)}\left(x_j(t)+\sum_{i=1}^N \frac{A_i(t)-A_i(s)}{\alpha_i(s)} V_{i j}(t)\right) .
$$

The state variables $x_i, V_{i j}$ form a joint Markov process and admit the differential representations:
$$
\begin{aligned}
d x_i(t) & =\left(x_i(t) \frac{d}{d t} \ln \left(\alpha_i(t)\right)+\sum_{j=1}^N V_{i j}(t)\right) d t+\beta_i(t) d W(t), \\
d V_{i j}(t) & =\left(\beta_i(t) \beta_j(t)+V_{i j}(t) \frac{d}{d t}\left(\ln \left(\alpha_i(t) \alpha_j(t)\right)\right)\right) d t .
\end{aligned}
$$

In particular, we obtain:
$$
r(t)=f(0, T)+\sum_{j=1}^N x_j(t),
$$
$$
\begin{aligned}
P(t, T)= & \frac{P(0, T)}{P(0, t)} \exp \left(-\sum_{i=1}^N \frac{A_i(T)-A_i(t)}{\alpha_i(t)} x_i(t)\right) \\
& \exp \left(-\sum_{i, j=1}^N \frac{\left(A_i(T)-A_i(t)\right)\left(A_j(T)-A_j(t)\right)}{2 \alpha_i(t) \alpha_j(t)} V_{i j}(t)\right) .
\end{aligned}
$$
    
\end{theorem}

\begin{remark}
$$
P(t, T)=\exp \left(-\int_t^T f(t, s) d s\right)
$$

$$
f(0, t)=f^M(0, t)  \forall  t  \geq 0
$$

$$
P(0, T)=P^M(0, T)  \forall  T \geq 0 .
$$

$$
\exp \left(-\int_t^T f(t, s) d s\right)=P(t, T)=\mathbb{E}_{\mathbb{Q}}\left(\exp \left(-\int_t^T r(s) ds\right)\right)
$$

$$
d f(t, T)=\mu_f(t, T) d t+\sigma_f(t, T) d W(t)
$$
for a $d$-dimensional Brownian motion $W$ (.) and suitable stochastic processes $\mu_f, \sigma_f$, then we must have,
$$
\mu_f(t, T)=\sigma_f(t, T) \int_t^T \sigma_f(t, s) d s \quad \text {drift condition, under $\mathbb{Q}$ }
$$ 

\end{remark}

\section*{Concepts of stochastic investment models: The Vasiček model}

\begin{proof} 

The Vasiček \footnote{\href{https://quantgirluk.github.io/Understanding-Quantitative-Finance/AOS.html}{Understanding-Quantitative-Finance}}\footnote{\href{https://hannoreuvers.github.io/post/vasicek/}{The Vasiček short-rate model}} model is an interest rate model which specifies the short rate $r(t)$ under the risk-neutral dynamics (or $\mathbb{Q}$-dynamics) as
$$
d r(t)=\kappa(\theta-r(t)) d t+\sigma d W(t),
$$

with initial condition $r(0)=r_0$ and $W(t)$ denoting a standard Brownian motion driving the stochastic differential equation. The Itô lemma implies:
$$
\begin{aligned}
d f(r(t), t) & =\kappa r(t) e^{\kappa t} d t+e^{\kappa t} d r(t) \\
& =\kappa r(t) e^{\kappa t} d t+e^{\kappa t}[\kappa(\theta-r(t)) d t+\sigma d W(t)] \\
& =\kappa \theta e^{\kappa t} d t+\sigma e^{\kappa t} d W(t) .
\end{aligned}
$$

$$
\begin{aligned}
f(r(t), t) & -f(r(0), 0)=r(t) e^{\kappa t}-r_0=\int_0^t d f(r(s), s) \\
& =\int_0^t \kappa \theta e^{\kappa s} d s+\int_0^t \sigma e^{\kappa s} d W(s) .
\end{aligned}
$$

$$
r(t)=r_0 e^{-\kappa t}+e^{-\kappa t} \int_0^t \kappa \theta e^{\kappa s} d s+e^{-\kappa t} \int_0^t \sigma e^{\kappa s} d W(s) .
$$

$$
\begin{aligned}
\mathbb{E}[r(t)] & =r_0 e^{-\kappa t}+\theta e^{-\kappa t}\left[e^{\kappa t}-1\right]=r_0 e^{-\kappa t}+\theta\left[1-e^{-\kappa t}\right] \\
& =r_0 e^{-\kappa t}+\theta \kappa \Lambda(t)
\end{aligned}
$$

$$
\operatorname{Var}[r(t)]=\sigma^2 e^{-2 \kappa t} \int_0^t e^{2 \kappa s} d s=\frac{\sigma^2}{2 \kappa}\left[1-e^{-2 \kappa t}\right]=\frac{1}{2} \sigma^2 \Lambda(2 t) .
$$

$$
\begin{aligned}
& f(r(t), t)-f(r(t-h), t-h)=\int_{t-h}^t d f(r(s), s) \\
& \quad=\theta\left(e^{\kappa t}-e^{\kappa(t-h)}\right)+\int_{t-h}^t \sigma e^{\kappa s} d W(s) \\
& \Longleftrightarrow r(t)=\theta\left(1-e^{-\kappa h}\right)+e^{-\kappa h} r(t-h)+e^{-\kappa t} \int_{t-h}^t \sigma e^{\kappa s} d W(s) .
\end{aligned}
$$

$$
\begin{aligned}
& \operatorname{Var}\left[e^{-\kappa t} \int_{t-h}^t \sigma e^{\kappa s} d W(s)\right] \\
& \quad=\sigma^2 e^{-2 \kappa t} \int_{t-h}^t e^{2 \kappa s} d s=\frac{\sigma^2}{2 \kappa}\left[1-e^{-2 \kappa h}\right]=\sigma^2 h\left[\frac{e^{-2 \kappa h}-1}{-2 \kappa h}\right] \\
& \quad=: \sigma^2 h \alpha(-2 \kappa h),
\end{aligned}
$$

\end{proof}


\begin{theorem}
Bond and option prices in the Vasiček model. In the Vasiček model we have:
(a) T-zero bond prices of the form:
$$
P(t, T)=e^{-B(t, T) r(t)+A(t, T)}
$$

with $A$ and $B$ given by,
$$
\begin{aligned}
& B(t, T)=\frac{1}{\kappa}\left(1-e^{-\kappa(T-t)}\right), \\
& A(t, T)=\left(\theta-\frac{\sigma^2}{2 \kappa^2}\right)(B(t, T)-T+t)-\frac{\sigma^2}{4 \kappa} B(t, T) .
\end{aligned}
$$
(b) Bond call and put option prices of the form,
$$
\begin{aligned}
\operatorname{Call}(t, T, S, K) & =P(t, S) \Phi\left(d_1(t)\right)-K P(t, T) \Phi\left(d_2(t)\right), \\
P u t(t, T, S, K) & =K P(t, T) \Phi\left(-d_2(t)\right)-P(t, S) \Phi\left(-d_1(t)\right)
\end{aligned}
$$
with
$$
d_{1 / 2}(t)=\frac{\ln \left(\frac{P(t, S)}{P(t, T) K}\right) \pm \frac{1}{2} \bar{\sigma}^2(t)}{\bar{\sigma}(t)}, \bar{\sigma}(t)=\sigma \sqrt{\frac{1-e^{-2 \kappa(T-t)}}{2 \kappa}} B(T, S)
$$
where $K$ denotes the strike and $T$ the maturity of the options, and $S \geq T$ is the maturity of the underlying zero bond.
(c) Prices for caps with face value $V$, level $L$, payment times $t_1<\ldots<t_n$
$$
\begin{aligned}
& \operatorname{Cap}(t ; V, L, \sigma)= \\
& \quad V \sum_{i=1}^n\left(P\left(t, t_{i-1}\right) \Phi\left(\tilde{d}_{1, i}(t)\right)-\left(1+\delta_i L\right) P\left(t, t_i\right) \Phi\left(\tilde{d}_{2, i}(t)\right)\right)
\end{aligned}
$$
for $t<t_0<t_1$ with
$$
\begin{gathered}
\tilde{d}_{1 / 2, i}(t)=\frac{1}{\bar{\sigma}_i(t)} \ln \left(\frac{P\left(t, t_{i-1}\right)}{\left(1+\delta_i L\right) P\left(t, t_i\right)}\right) \pm \frac{1}{2} \bar{\sigma}_i(t), \\
\bar{\sigma}_i(t)=\sigma \sqrt{\frac{1-e^{-2 \kappa\left(t_{i-1}-t\right)}}{2 \kappa}} B\left(t_{i-1}, t_i\right), \quad \delta_i=t_i-t_{i-1} .
\end{gathered}
$$

\end{theorem}


\section*{Concepts of stochastic investment models: The Longstaff-Schwartz model}

\vspace{3mm}

 $\left(\Omega, \mathcal{F},\left(\mathcal{F}_t\right), \mathbb{P}\right)$ a filtered probability space
 Let $X$ denote the price of a risky asset whose dynamics follow a stochastic process $X=\{X(t): 0 \leq t \leq T\}$. Let $B$ denote the process representing the money market account, i.e. $B=\{B(t): 0 \leq t \leq T\}$, where
$$
B(t)=\exp \left(\int_0^t r_u d u\right),
$$

and $r$ is the instantaneous short rate process.
Consider an American put option for the underlying $X$ with strike $K$ and maturity $T$.

At maturity $T$ the value of the option equals the payoff:
$$
Y(T)=(K-X(T))^{+}=\max \{K-S(T), 0\} \text {. }
$$

At any time $t \in[0, T)$ the option buyer has two choices.

If the option seller knew in advance which stopping time $\tau_0$ the investor will use, then he would set the price as:

$$
\mathbb{E}_Q\left[\frac{Y\left(\tau_0\right)}{B\left(\tau_0\right)}\right] .
$$

However, the optimal stopping time is not know. Thus, the option seller has to prepare for the worst possible case and charge the maximum value, i.e:
$$
\sup _{\tau \in \mathcal{T}} \mathbb{E}_Q\left[\frac{Y(\tau)}{B(\tau)}\right],
$$

where $\mathcal{T}$ is the class of admisible stopping times taking values in $[0, T]$.

\begin{definition}[Longstaff-Schwartz]

Define $Z=\{Z(t): 0 \leq t \leq T\}$, given by:
$$
Z(t)=\sup _{\tau \in \mathcal{T}_{[t, T]}} \mathbb{E}_Q\left[\left.\frac{Y(\tau)}{B(\tau)} \right\rvert\, \mathcal{F}_t\right] B(t) .
$$

Then $Z(t) / B(t)$ is the smallest $\mathbb{Q}$-supermartingale satisfying $Z(t) \geq Y(t)$. 
$$
\tau(t)=\inf \{s \geq t: Z(s)=Y(s)\} .
$$

$$
\mathbb{E}_Q\left[\left.\frac{Y(\tau)}{B(\tau)} \right\rvert\, \mathcal{F}_t\right]=\sup _{\tau \in \mathcal{T}_{[t, \tau]}} \mathbb{E}_Q\left[\left.\frac{Y(\tau)}{B(\tau)} \right\rvert\, \mathcal{F}_t\right] 
$$

\end{definition}

\section*{The Girsanov theorem: Radon-Nikodym derivative}

$$
\mathbb{P}^*(A)=\int_A \rho_t(\omega) d \mathbb{P}(\omega), \quad A \in \mathcal{F}_t
$$

$$
\left.\frac{d \mathbb{P}^*}{d \mathbb{P}}\right|_{\mathcal{F}_t}=\rho_t
$$

The process $\rho_t$ is called the Radon-Nikodym derivative \footnote{\href{ https://cel.hal.science/cel-00398075}{Valuation and Hedging of Credit Derivatives}}) of $\mathbb{P}^*$ with respect to $\mathbb{P}$ restricted to $\mathcal{F}_t$.

$$
\mathbb{E}^*[X]=\int_{\Omega} X(\omega) d \mathbb{P}^*(\omega)=\int_{\Omega} X(\omega) \frac{d \mathbb{P}^*}{d \mathbb{P}}(\omega) d \mathbb{P}(\omega)=\mathbb{E}\left[X \frac{d \mathbb{P}^*}{d \mathbb{P}}\right]
$$

where $\mathbb{E}^*$ and $\mathbb{E}$ denote expected values with respect to the probability measures $\mathbb{P}^*$ and $\mathbb{P}$, 

$$
\mathbb{E}^*\left[X \mid \mathcal{F}_t\right]=\frac{\mathbb{E}\left[\left.X \frac{d \mathbb{P}^*}{d \mathbb{P}} \right\rvert\, \mathcal{F}_t\right]}{\rho_t}
$$

\begin{theorem}

Consider the stochastic differential equation, with Lipschitz coefficients,
$$
d X_t(\omega)=f\left(X_t(\omega)\right) d t+\sigma\left(X_t(\omega)\right) d W_t(\omega), \quad x_0
$$ under $\mathbb{P}$. Let be given a new drift $f^*(x)$ and assume $\left(f^*(x)-f(x)\right) / \sigma(x)$ to be bounded. Define the measure $\mathbb{P}^*$ by, 

$$
\begin{aligned}
\left.\frac{d \mathbb{P}^*}{d \mathbb{P}^2}(\omega)\right|_{\mathcal{F}_t}=\exp \{ & -\frac{1}{2} \int_0^t\left(\frac{f^*\left(X_s(\omega)\right)-f\left(X_s(\omega)\right)}{\sigma\left(X_s(\omega)\right)}\right)^2 d s \\
& \left.+\int_0^t \frac{f^*\left(X_s(\omega)\right)-f\left(X_s(\omega)\right)}{\sigma\left(X_s(\omega)\right)} d W_s(\omega)\right\}
\end{aligned}
$$

Then $\mathbb{P}^*$ is equivalent to $\mathbb{P}$. Moreover, the process $W^*$ defined by,

$$
d W_t^*(\omega)=-\left[\frac{f^*\left(X_t(\omega)\right)-f\left(X_t(\omega)\right)}{\sigma\left(X_t(\omega)\right)}\right] d t+d W_t(\omega)
$$

is a Brownian motion under $\mathbb{P}^*$,

$$
d X_t(\omega)=f^*\left(X_t(\omega)\right) d t+\sigma\left(X_t(\omega)\right) d W_t^*(\omega), \quad x_0
$$

$$
d S_t(\omega)=\mu S_t(\omega) d t+\sigma S_t(\omega) d W_t(\omega)
$$

$$
d S_t(\omega)=r S_t(\omega) d t+\sigma S_t(\omega) d W_t^*(\omega)
$$

$$
\left.\frac{d \mathbb{P}^*}{d \mathbb{P}^2}(\omega)\right|_{\mathcal{F}_t}=\exp \left\{-\frac{1}{2}\left(\frac{\mu-r}{\sigma}\right)^2 t-\frac{\mu-r}{\sigma} W_t(\omega)\right\}
$$.

\end{theorem}

\section*{Bayesian optimization of various stochastic investment models using the “GPyOpt” framework}

Bayesian model formulation:\\

$$
\begin{gathered}
\theta \sim \pi(\theta) \quad \text {a priori law} \\
Y_i \mid \theta \stackrel{i i d}{\sim} f(y \mid \theta) \quad \text {sampling model}
\end{gathered}
$$

$$
p(\theta \mid \boldsymbol{y})=\frac{f(\boldsymbol{y} \mid \theta) \pi(\theta)}{f(\boldsymbol{y})}
$$

where $p(\theta\mid\boldsymbol{y})$ is the a posteriori distribution, $f(\boldsymbol{y} \mid \theta)$ is the likelihood (inherited from the sampling model), $\pi(\theta)$ is the a priori distribution of the parameters $\theta$ and $f(\boldsymbol{y})=\int f(\boldsymbol{y}\mid\theta) \pi(\theta)$ is the marginal distribution of the data, i.e. the constant (relative to $\theta$) of standardization.\\

Obtaining the law a posteriori:

$$
p(\theta \mid \boldsymbol{y}) \propto f(\boldsymbol{y} \mid \theta) \pi(\theta)
$$

 Jeffreys' weakly informative a priori law:

$$
\pi(\theta) \propto \sqrt{I(\theta)} 
$$

Predictive distribution:

$$
f_{Y_{n+1}}(y \mid \boldsymbol{y})=\int f_{Y_{n+1}}(y \mid \theta) p(\theta \mid \boldsymbol{y}) \mathrm{d} \theta
$$ .

\begin{figure}[H]
\centering
\includegraphics[width=0.46\textwidth]{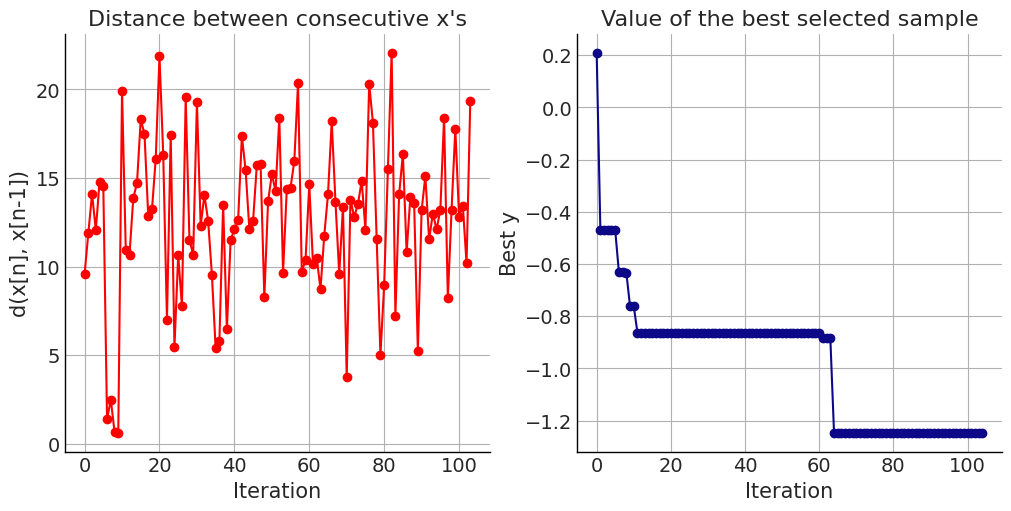}
\caption{Bayesian optimization.}
\label{fig:Bayesian_optimization}
\end{figure}

Bayesian optimisation: Global optimization \footnote{\href{https://sheffieldml.github.io/GPyOpt/}{GPyOpt}} with different acquisition    \footnote{\href{https://www.blopig.com/blog/wp-content/uploads/2019/10/GPyOpt-Tutorial1.html}{acquisition functions}} functions \cite{hooker2022use}, \cite{roussel2024bayesian}, \cite{khatamsaz2023physics}
Formally, the purpose of BO (Bayesian-Optimization) \cite{ekstrom2019bayesian},\cite{shields2021bayesian} is to retrieve the optimum $\mathbf{x}^{\star}$ of a black-box function (Figure \ref{fig:Bayesian_optimization}) $f(\mathbf{x})$ where $\mathbf{x} \in \mathcal{X}$ and $\mathcal{X}$ is the input space where $f(\mathbf{x})$ can be observed. We want to retrieve $\mathbf{x}^{\star}$ such that,
$$
\mathbf{x}^{\star}=\arg \min _{\mathbf{x} \in \mathcal{X}} f(\mathbf{x}),
$$
 We can define a BO method by, 
$$
\mathcal{A}=(\mathcal{M}, \alpha(\cdot), p(f(\mathbf{x}) \mid \mathcal{D})), 
$$ (optimization results can be viewed on ART \footnote{\href{https://github.com/Trusted-AI/adversarial-robustness-toolbox/pull/2443}{ART-IBM}}, readers can modify the Bayesian optimization parameters as they wish, but the results will always be the same.)

\begin{algorithm}[ht]
\DontPrintSemicolon
\SetAlgoLined
\KwIn{Objective function $f(x)$, Bounds $B$}
\KwOut{Optimized parameters $x^*$, Best value $v^*$}

Initialize Bayesian optimizer with $f$ and $B$\;
\While{Stopping criterion not met}{
    choose initial $\theta^{(1)}, \theta^{(2)}, \ldots \theta^{(k)}$, where $k \geq 2$\;
     evaluate the objective function $f(\theta)$ to obtain $y^{(i)}=f\left(\theta^{(i)}\right)$ for $i=1, \ldots, k$ \;
     initialize a data vector $\mathcal{D}_k=\left\{\left(\theta^{(i)}, y^{(i)}\right)\right\}_{i=1}^k$\;
     select a statistical model for $f(\theta)$  \\
  for $\{n=k+1, k+2, \ldots\}$\;
     select $\theta^{(n)}$ by optimizing (maximizing) the acquisition function\;
      $\theta^{(n)}=\underset{\theta}{\arg \max } \mathcal{A}\left(\theta \mid \mathcal{D}_{n-1}\right)$\;
      evaluate the objective function to obtain $y^{(n)}=f\left(\theta^{(n)}\right)$ \;
      augment the data vector $\mathcal{D}_n=\left\{\mathcal{D}_{n-1},\left(\theta^{(n)}, y^{(n)}\right)\right\}$\;
        update the statistical model for $f(\theta)$ \;
}
Return $x^*$ and $v^*$\;

\caption{Bayesian Optimization Process}
\label{alg:bayesianOptimization}
\end{algorithm}

Acquisition functions:
\begin{enumerate}
\item  Expected Improvement (EI)
\item  Lower Confidence Bound (LCB)
\end{enumerate}

We adopt the notation below and write $\mathcal{A}(\theta) \equiv \mathcal{A}(\theta \mid D)$.

$$
\begin{aligned}
\mathcal{A}_{\mathrm{EI}}(\theta)=\left\langle\max \left(0, f_{\min }-f(\theta)\right)\right\rangle & \\ =\int_{-\infty}^{\infty} \max \left(0, f_{\min }-f\right) \mathcal{N}\left(f(\theta) \mid \mu(\theta), \sigma(\theta)^2\right) d f(\theta) \\
& \\ =\int_{-\infty}^{f_{\min }}\left(f_{\min }-f\right) \frac{1}{\sqrt{2 \pi \sigma^2}} \exp \left[-\frac{(f-\mu)^2}{2 \sigma^2}\right] d f \\
& \\ =\left(f_{\min }-\mu\right) \Phi\left(\frac{f_{\min }-\mu}{\sigma}\right)+\sigma \phi\left(\frac{f_{\min }-\mu}{\sigma}\right) \\
& \\ =\sigma(z \Phi(z)+\phi(z))
\end{aligned}
$$

where

$$
\mathcal{N}\left(f(\theta) \mid \mu(\theta), \sigma(\theta)^2\right) 
$$

$$
\mathcal{A}(\theta)_{\mathrm{LCB}}=\beta \sigma(\theta)-\mu(\theta)
$$

\bibliographystyle{IEEEtran}

\bibliography{IEEEabrv,refs}

\end{document}